\documentclass[aps,twocolumn,showpacs,superscriptaddress]{revtex4}

\usepackage{graphicx}% Include figure files
\usepackage{bm}% bold math
\usepackage{amsmath}% \text{}
\usepackage{dcolumn}%

\newcommand{\eg}{\ensuremath{e_{g}}}
\newcommand{\tg}{\ensuremath{t_{2g}}}
\newcommand{\dxy}{\ensuremath{d_{xy}}}
\newcommand{\dxz}{\ensuremath{d_{xz}}}
\newcommand{\dyz}{\ensuremath{d_{yz}}}

\newcommand{\mb}{\ensuremath{\mu_{\text{B}}}}

\newcolumntype{/}{D{/}{/}{2,2}}  % / as a delimiter
\newcolumntype{.}{D{.}{.}{0}}  % / as a delimiter

\begin{document}

\title{Density functional theory of resonant inelastic x-ray scattering in the
  quasi-one-dimensional dimer iridate Ba$_3$InIr$_2$O$_9$ }

\author{D.A. Kukusta}

\affiliation{G. V. Kurdyumov Institute for Metal Physics of the
  N.A.S. of Ukraine, 36 Academician Vernadsky Boulevard, UA-03142
  Kyiv, Ukraine}

\author{L.V. Bekenov}

\affiliation{G. V. Kurdyumov Institute for Metal Physics of the
  N.A.S. of Ukraine, 36 Academician Vernadsky Boulevard, UA-03142
  Kyiv, Ukraine}

\author{V.N. Antonov}

\affiliation{G. V. Kurdyumov Institute for Metal Physics of the
  N.A.S. of Ukraine, 36 Academician Vernadsky Boulevard, UA-03142
  Kyiv, Ukraine}

\affiliation{Max-Planck-Institute for Solid State Research,
  Heisenbergstrasse 1, 70569 Stuttgart, Germany}

\date{\today}

\begin{abstract}

We have investigated the electronic structure of Ba$_3$InIr$_2$O$_9$
within the density-functional theory (DFT) using the generalized
gradient approximation while considering strong Coulomb correlations
(GGA+$U$) in the framework of the fully relativistic spin-polarized
Dirac linear muffin-tin orbital band-structure method. We have
investigated resonant inelastic x-ray scattering (RIXS) spectra at the
Ir $L_3$ $K$ edge. The calculated results are in good agreement with
experimental data. The RIXS spectrum of Ba$_3$InIr$_2$O$_9$ at the Ir
$L_3$ edge possesses several sharp features below 2 eV corresponding
to transitions within the Ir {\tg} levels. The excitation located from
2 to 5 eV is due to {\tg} $\rightarrow$ {\eg} transitions. The third
wide structure situated at 5$-$12 eV appears due to charge transfer
transitions. We have also presented comprehensive theoretical
calculations of the RIXS spectrum at the oxygen $K$ edge.

\end{abstract}

\pacs{75.50.Cc, 71.20.Lp, 71.15.Rf}

\maketitle

\section{Introduction}

\label{sec:introd}

In 5$d$ transition metal compounds the energy scale of the spin-orbit coupling
(SOC) is comparable with the on-site Coulomb interaction and the crystal-field
energy. Due to the strong competition between these interactions fascinating
electronic states can arise. SOC in such systems splits the {\tg} orbitals
into a quartet ($J_{\rm{eff}}$ = $\frac{3}{2}$) and a doublet ($J_{\rm{eff}}$
= $\frac{1}{2}$) \cite{JaKh09,CPB10,WCK+14}. In 5$d^5$ (Ir$^{4+}$) iridium
oxides, such as Sr$_2$IrO$_4$, the quartet $J_{\rm{eff}}$ = $\frac{3}{2}$ is
fully occupied, and the relatively narrow $J_{\rm{eff}}$ = $\frac{1}{2}$
doublet, occupied by one electron, can be splitted by moderate Hubbard
$U_{\rm{eff}}$ with opening a small band gap called the relativistic Mott gap
\cite{KJM+08,MAV+11,AUU18,AKB24}. Iridates have been at the center of an
intensive research in recent years for novel phenomena, such as topological
insulators \cite{QZ10,Ando13,WBB14,BLD16}, Mott insulators
\cite{KJM+08,KOK+09,WSY10,MAV+11}, Weyl semimetals \cite{WiKi12,GWJ12,SHJ+15},
and quantum spin liquids (QSLs) \cite{KAV14,Bal10,SaBa17}.

The specific features of the systems with spin and orbital degeneracy strongly
depend on the local geometry. The most typical cases, widely discussed in the
literature, are those with MO$_6$ octahedra (M is a transition-metal ion)
sharing a common oxygen (or common corner), such as the LaMnO$_3$ perovskite or
layered systems as La$_2$CuO$_4$, and the situation with two common oxygen
for neighboring octahedra (octahedra with a common edge), met in many layered
systems with triangular lattices such as NaCoO$_2$ and LiNiO$_2$. The features
of spin-orbital systems in both these cases were studied in detail; see, e.g.,
Ref.  \cite{book:Khomski14}.  However, there exists yet the third typical
geometry, which is also very often met in many real materials: the case of
octahedra with a common face (three common oxygens). This case has relatively
small attention in the literature. Actually, there are many transition-metal
compounds with face-sharing geometry \cite{KKS+15}. Such materials include,
for example, hexagonal crystals such as BaCoO$_3$ \cite{YZA+99}, BaVS$_3$
\cite{GVW69}, or CsCuCl$_3$ \cite{Hir77}.  Many other similar systems have
finite face-sharing blocks, e.g., BaIrO$_3$ \cite{SiCh91}, BaRuO$_3$
\cite{HoSl97,ZYY+07}, or Ba$_4$Ru$_3$O$_{10}$ \cite{CDF+00,KRD+11} with blocks
of three such face-sharing octahedra, connected between themselves by common
corners; or blocks of two such octahedra as in large series of systems with
the general formula A$_3$BM$_2$O$_9$
\cite{KR77,KSF+12,SAS+13,FGG80,SKA+13,RHE+99,Strel13,TKE82}, where A is Ba,
Ca, Sr, Li, or Na, and the face-sharing MO$_6$ octahedra of transition metals are
separated by BO$_6$ octahedra (which have common corners with MO$_6$). Such
systems have very diverse properties: some of them are metallic \cite{RJZ+99};
others are insulators \cite{KR77} or undergo a metal-insulator transition
\cite{KSF+12}. Materials in this family with magnetic B sites have been shown
to exhibit quantum spin-liquid behavior, in particular Ba$_3$NiSb$_2$O$_9$
\cite{CLB+11,QBM+16,FBC+17} and Ba$_3$IrTi$_2$O$_9$ \cite{DMK+12}, while
others, Ba$_3$CuSb$_2$O$_9$ \cite{ZCL+11,NKK+12,QBK+12} and
Ba$_3$ZnIr$_2$O$_9$ \cite{NMB+16}, exhibit possible quantum spin-orbital
liquids. Furthermore, Ba$_3$CoSb$_2$O$_9$ has allowed for some of the first
studies on the magnetization process of a truly triangular spin-$\frac{1}{2}$
antiferromagnet \cite{STM+12,ZXH+12,SKT+13,KZK+15,QLP+15}.  

The series of perovskites Ba$_3$MIr$_2$O$_9$ (M = an alkaline-earth, transition
metal, or rare-earth element) is rather interesting also because a wide range
of elements with a variety of Ir oxidation states can be fitted by changing
component M. The nominal oxidation state of Ir can be 5.5+ (for M = Li$^+$,
Na$^+$ \cite{KSD+04}), 5+ (for M = Zn$^{2+}$, Mg$^{2+}$, Ca$^{2+}$, Cd$^{2+}$
\cite{SDH06}), 4.5+ (for M=Y$^{3+}$, Sc$^{3+}$, In$^{3+}$, Lu$^{3+}$
\cite{SDH06,DoHi04}), and 4+ (for M=Ti$^{4+}$, Zr$^{4+}$ \cite{SDH06}). There
are several experimental studies on the 6H Ba$_3$MIr$_2$O$_9$ compounds with
M= Zn, Mg, Ca, Sr, Y, Sc, and In
\cite{NMB+16,SDH06,DMK+13,NBB+18,DMO+17,KBN+19}, which show diverse
ground-state properties. For instance, Ba$_3$ZnIr$_2$O$_9$ was reported to
exhibit a spin-orbital liquid state, with an effective moment of $\sim$0.26
{\mb}/Ir site \cite{NMB+16}.  The closely related Ba$_3$CdIr$_2$O$_9$
\cite{KBN+19} and Ba$_3$MgIr$_2$O$_9$ \cite{NBB+18} were also seen to exhibit
no long-range magnetic order down to the lowest measured temperatures.
However, Ba$_3$CaIr$_2$O$_9$ and Ba$_3$SrIr$_2$O$_9$ systems were observed to
exhibit weak-dimer-like and ferromagnetic (FM)-like features in the magnetic
susceptibility, respectively, though these transitions could not be identified
in heat capacity measurements \cite{NBB+18}. 

The materials with a fractional oxidation state of Ir are especially very
interesting because the fractional oxidation state of Ir (accompanied by a
unique crystallographic site) can lead them to fascinating ground states.
Such materials have been less explored theoretically, despite the fact that
their experimental magnetic response reveals several peculiarities. There are
three holes per dimer in Ba$_3$M$^{3+}$Ir$_2^{4.5+}$O$_9$ compounds. At high
temperatures, where the mixed-valence dimers can be seen as isolated, the magnetic
susceptibility deviates from the conventional Curie-Weiss behavior
\cite{DMH02,SDH06,NaRa17} suggesting nontrivial temperature evolution of the
local magnetic moment. At low temperatures, interactions between the dimers
become important, and signatures of frustrated magnetic behavior \cite{ZAS+17}
including possible formation of a spin-liquid ground state have been reported
\cite{DMO+17}. The linear fit to the high-temperature part produces the
effective moments of 1.53 {\mb}/f.u. (for M = Ir) and 1.79 {\mb}/f.u. (for M =
Sc) that were interpreted as the $S$ = 1 state of the mixed-valence dimer
\cite{SDH06}.  At lower temperatures, the interactions between the dimers
occur.  Ba$_3$YIr$_2$O$_9$ undergoes long-range magnetic ordering at 4.5 K
\cite{DKM+14,PBL+15} but Ba$_3$ScIr$_2$O$_9$ does not show no long-range magnetic
order down to at least 2 K \cite{DKM+14}.  At the same time,
according to thermodynamic measurements, nuclear magnetic resonance, and muon spin 
resonance, Ba$_3$In$^{3+}$Ir$_2^{4.5+}$O$_9$ shows persistent spin dynamics down to 20 mK
without any long-range magnetic order \cite{DMO+17}. This can point toward 
a possible gapless spin-liquid ground state in this perovskite.

Here we report a theoretical investigation of the electronic and magnetic
structures of the Ba$_3$InIr$_2$O$_9$ perovskite. The structure of this compound
consists of InO$_6$ and IrO$_6$ octahedra. The latter octahedra share a face
and develop along the crystallographic $c$ axis so called Ir$_2$O$_9$ dimers,
which are connected through corners with the InO$_6$ octahedra and form an
edge-shared triangular lattice in the $ab$ plane. Therefore, such a compound would
become geometrically frustrated in the presence of the antiferromagnetic (AFM)
interaction. Novel properties are expected from this structural arrangement in
addition to those driven by SOC.

In this article, we focus our attention on the RIXS properties of
Ba$_3$InIr$_2$O$_9$.  Since the first publication by Kao {\it et al.} on NiO
\cite{KCH+96}, the RIXS method has shown remarkable progress as a
spectroscopic technique to record the momentum and energy dependence of
inelastically scattered photons in complex materials. RIXS rapidly became the
forefront of experimental photon science \cite{AVD+11,GHE+24}. RIXS combines
spectroscopy and inelastic scattering to probe the electronic structure of
materials. This method is an element- and orbital-selective X-ray
spectroscopy technique, based on a two-step, two-photon resonant process. It
combines X-ray emission spectroscopy (XES) with X-ray absorption spectroscopy
(XAS) by measuring the coherent X-ray emission at incident X-ray photon
energy within the near edge X-ray absorption spectrum. In the first step
(X-ray absorption), an electron of the absorbing atom is resonantly excited
from a core level to an empty state. The resulting state, called the
intermediate state, carries a core hole with a very small lifetime. In the
second step (X-ray emission), the system radiatively decays into a final state
in which the core hole is filled by another electron accompanied by
photon-out emission. The polarization of the incoming and outgoing light and
the resonant energy are involved in the RIXS process, making RIXS a
simultaneous spectroscopy and scattering technique. RIXS has a number of
unique features in comparison with other spectroscopic techniques. It covers a
large scattering phase space and requires only small sample volumes. It is
also bulk sensitive, polarization dependent, as well as element and orbital
specific \cite{AVD+11}. A detail comparison with other spectroscopic techniques
can be found in recent review article \cite{GHE+24}. Spectral broadening owing
to a short core hole lifetime can be reduced to produce the RIXS spectra with high
resolution. It permits direct measurements of phonons, plasmons,
single-magnon, and orbitons as well as other many-body excitations in strongly
correlated systems, such as cuprates, nickelates, osmates, ruthenates, and
iridates, with complex low-energy physics and exotic phenomena in the energy and
momentum space.

There is great progress in the RIXS experiments over the past
decade. The most calculations of the RIXS spectra of various materials
have been carried out using the atomic multiplet approach with some
adjustable parameters and the number of theoretical first-principle
calculations of RIXS spectra is extremely limited. In this paper, we
report a theoretical investigation from the first principles of the
RIXS spectra of Ba$_3$InIrO$_9$. Recently, the RIXS measurements have
been successfully performed at the Ir $L_3$ edge in Ba$_3$InIrO$_9$ by
Revelli {\it et al.} \cite{RSM+22} in the energy range up to 5.5
eV. In addition to the elastic peak centered at zero energy loss, the
spectrum consists of several peaks below 2 eV with a high energy
shoulder at $\sim$1.8 eV and a peak at 4 eV.

We carry out here a detailed study of the electronic structure and
RIXS spectra of Ba$_3$InIrO$_9$ in terms of the density functional
theory. Our study sheds light on the important role of band structure
effects and transition metal 5$d$ $-$ oxygen 2$p$ hybridization in the
spectral properties of 5$d$ oxides. We use the {\it ab initio}
approach using the fully relativistic spin-polarized Dirac linear
muffin-tin orbital band-structure method. Both the generalized
gradient approximation (GGA) and the GGA+$U$ approach are used to
assess the sensitivity of the RIXS results to different treatment of
the correlated electrons.

The paper is organized as follows. The crystal structure of Ba$_3$InIrO$_9$
and computational details are presented in Sec. II. Section III presents the
electronic and magnetic structures of Ba$_3$InIrO$_9$. In Sec. IV, the
theoretical investigation of the RIXS spectrum of Ba$_3$InIrO$_9$ at the Ir
$L_3$ edge is presented, the theoretical results are compared with
experimental measurements. Finally, the results are summarized in Sec. V.

\section{Computational details}
\label{sec:details}

\subsection{RIXS}  

In the direct RIXS process \cite{AVD+11} the incoming photon with energy
$\hbar \omega_{\mathbf{k}}$, momentum $\hbar \mathbf{k}$, and polarization
$\bm{\epsilon}$ excites the solid from ground state $|{\rm g}\rangle$ with
energy $E_{\rm g}$ to intermediate state $|{\rm I}\rangle$ with energy
$E_{\rm I}$. During relaxation an outgoing photon with energy $\hbar
\omega_{\mathbf{k}'}$, momentum $\hbar \mathbf{k}'$ and polarization
$\bm{\epsilon}'$ is emitted, and the solid is in state $|f \rangle$ with
energy $E_{\rm f}$. As a result, an excitation with energy $\hbar \omega =
\hbar \omega_{\mathbf{k}} - \hbar \omega_{\mathbf{k}'}$ and momentum $\hbar
\mathbf{q}$ = $\hbar \mathbf{k} - \hbar \mathbf{k}'$ is created.  Our
implementation of the code for the calculation of the RIXS intensity uses
Dirac four-component basis functions \cite{NKA+83} in the perturbative
approach \cite{ASG97}. RIXS is the second-order process, and its intensity is
given by

\begin{eqnarray}
I(\omega, \mathbf{k}, \mathbf{k}', \bm{\epsilon}, \bm{\epsilon}')
&\propto&\sum_{\rm f}\left| \sum_{\rm I}{\langle{\rm
    f}|\hat{H}'_{\mathbf{k}'\bm{\epsilon}'}|{\rm I}\rangle \langle{\rm
    I}|\hat{H}'_{\mathbf{k}\bm{\epsilon}}|{\rm g}\rangle\over
  E_{\rm g}-E_{\rm I}} \right|^2 \nonumber \\ && \times
\delta(E_{\rm f}-E_{\rm g}-\hbar\omega),
\label{I1}
\end{eqnarray}
where the RIXS perturbation operator in the dipole approximation is given by
the lattice sum $\hat{H}'_{\mathbf{k}\bm{\epsilon}}=
\sum_\mathbf{R}\hat{\bm{\alpha}}\bm{\epsilon} \exp(-{\rm
  i}\mathbf{k}\mathbf{R})$, where $\bm{\alpha}$ are the Dirac matrices. The
sum over intermediate states $|{\rm I}\rangle$ includes the contributions
from different spin-split core states at the given absorption edge. The matrix
elements of the RIXS process in the frame of the fully relativistic Dirac LMTO
method were presented in Ref. \cite{AKB22a}.

\subsection{Crystal structure} 

The Rietveld refinement of room-temperature (300 K) neutron diffraction data for
Ba$_3$InIr$_2$O$_9$ obtained by Dey {\it et al.} \cite{DMO+17} reveals the
hexagonal crystal structure (space group $P6_3/mmc$, number 194). At 3.4 K the
symmetry is reduced to the monoclinic structure ($C2/c$ space group, number
15). The corresponding lattice parameters and Wyckoff positions are presented
in Table \ref{struc_tab_BIIO}. The mixed-valence Ba$_3$InIr$_2$O$_9$ shows a
high degree of structural order. Dey {\it et al.} \cite{DMO+17} were able to
obtain reasonable atomic displacement parameters in the fully ordered models
of both hexagonal and monoclinic structures, however, a weak site-mixing
scenario was observed with $\sim$2.8\% In$-$Ir intersite mixing.

\begin{figure}[tbp!]
\begin{center}
\includegraphics[width=0.90\columnwidth]{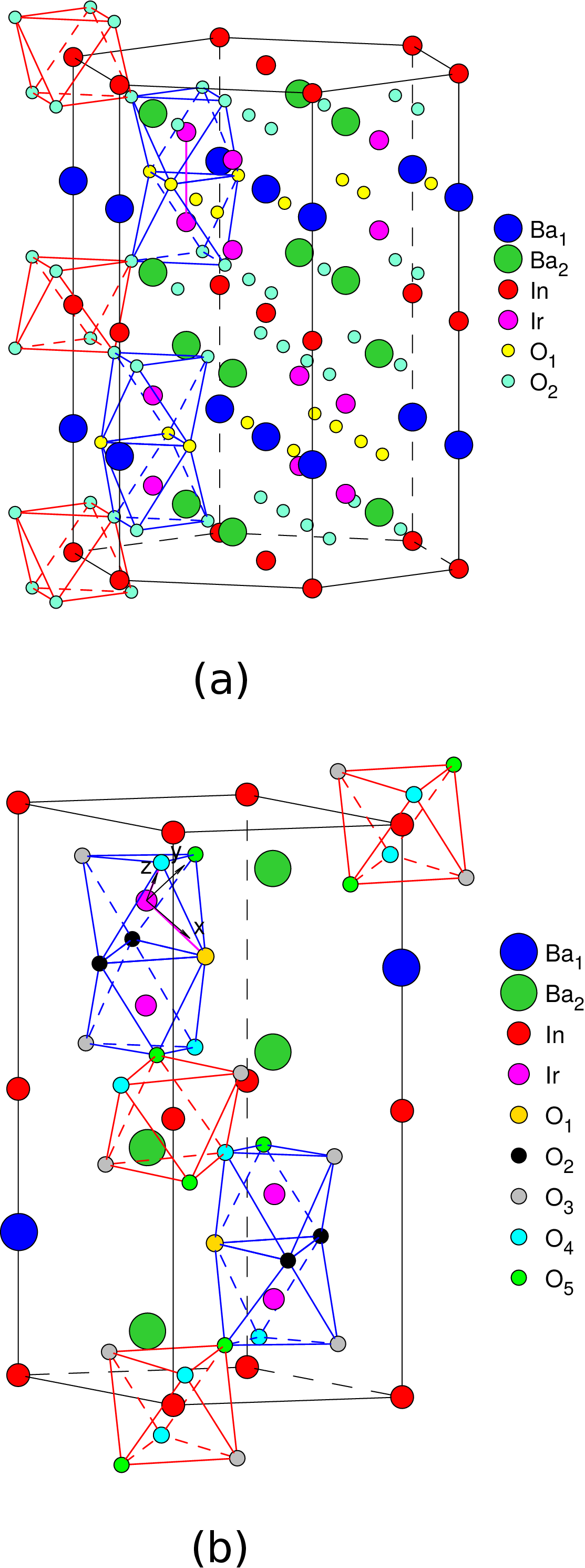}
\end{center}
\caption{\label{struc_BIIO}(Color online) The schematic representation of
  Ba$_3$InIr$_2$O$_9$ in the hexagonal $P6_3/mmc$ (a) and monoclinic $C2/c$ (b)
  crystal structures. }
\end{figure}

Both the hexagonal and monoclinic structures feature a single
crystallographic position of Ir, suggesting the intermediate-valence
Ir 4.5+ state. This is different from, e.g., Ba$_5$AlIr$_2$O$_{11}$,
where two sites of the Ir$_2$O$_9$ dimer belong to two different
crystallographic positions, thus making possible charge redistribution
within the dimer \cite{TWY+15}. The structure consists of Ir$_2$O$_9$
dimers, which are IrO$_6$ face-shared octahedra along the
crystallographic $c$ axis, connected through corners with InO$_6$
octahedra and forming an edge-shared triangular lattice in the $ab$
plane, with Ba ions sitting in typical for normal perovskites 12-fold
coordination sites (see Fig. \ref{struc_BIIO}). Ba$_3$InIr$_2$O$_9$
contains two formula units and hence two structural dimers in the unit
cell, which are connected via O-In-O paths along the $c$ axis.

%%%%%%%%%%%%%%%%%%%%%%%%%%%%
%%%%%%%%%%%%%%%%%%%%%%%%

The Ir$-$O$_1$ and Ir$-$O$_2$ interatomic distances for the hexagonal
$P6_3/mmc$ structure are equal to 2.034 and 1.962 \AA,
respectively. The corresponding Ir$-$O$_1$, Ir$-$O$_2$, Ir$-$O$_3$,
Ir$-$O$_4$, and Ir$-$O$_5$ distances for the monoclinic $C2/c$
structure are equal to 2.056, 2.023, 1.951, 2.017, and 1.958 \AA,
respectively.  The In$-$O$_2$ interatomic distance in InO$_6$
octahedra for the hexagonal $P6_3/mmc$ structure is equal to 2.118
\AA. The corresponding In$-$O$_3$, In$-$O$_4$, and In$-$O$_5$
distances for the monoclinic $C2/c$ structure are equal to 2.132,
2.105, and 2.101 \AA, respectively. The low-temperature monoclinic
distortion is primarily related to the tilting of IrO$_6$ and InO$_6$
octahedra. It has nearly no effect on relative positions of Ir
atoms. For example, the Ir-Ir distance within the dimer shrinks from
2.637 \AA\, at 300 K to 2.596\AA\, at 3.4 K, presumably, due to
thermal contraction \cite{DMO+17}.

%%%%%%%%%%%%%%%%%%%%%%%%%%%%%%%%%
\begin{table}[tbp!]
  \caption {The Wyckoff positions (WP) for the hexagonal $P6_3/mmc$
    and monoclinic $C2/c$ crystal structures of Ba$_3$InIr$_2$O$_9$
    (lattice constants $a$ = 5.8316 \AA, $c$ = 14.4877 \AA\, for the
    $P6_3/mmc$ structure and $a$ = 5.8152 \AA, $b$ = 10.0680 \AA, $c$
    = 14.4619 \AA, and $\beta$ = 90.854$^{\circ}$ for the $C2/c$
    structure) \cite{DMO+17}. }
\label{struc_tab_BIIO}
\begin{center}
\begin{tabular}{|c|c|c|c|c|c|}
\hline
Structure      & WP & Atom     & $x$      & $y$    & $z$     \\
\hline
               & $2b$   & Ba$_1$    &  0        & 0      &  0.25 \\
               & $4f$   & Ba$_2$    &  0.3333   & 0.6667 &  0.0889 \\
               & $2a$   & In        &  0        & 0      &  0      \\
$P6_3/mmc$     & $4f$   & Ir        &  0.3333   & 0.6667 &  0.1590 \\
               & $6h$   & O$_1$     &  0.4867   & 0.5133 &  0.25   \\
               & $12k$  & O$_2$     &  0.1715   & 0.3430 &  0.0841  \\
\hline
               & $4e$   & Ba$_1$    &  0        & 0.0010 &  0.25 \\
               & $8f$   & Ba$_2$    &  0.0049   & 0.3360 &  0.0889 \\
               & $4a$   & In        &  0        & 0      &  0      \\
               & $8f$   & Ir        & $-$0.0081 & 0.3334 &  0.8397 \\
$C2/c$         & $8f$   & O$_1$     &  0        & 0.4918 &  0.75   \\
               & $8f$   & O$_2$     &  0.2300   & 0.2573 &  0.7541  \\
               & $8f$   & O$_3$     & $-$0.0180 & 0.1717 &  0.9140  \\
               & $8f$   & O$_4$     &  0.2260   & 0.4158 &  0.9251  \\
               & $8f$   & O$_5$     & $-$0.2630 & 0.4144 &  0.9079  \\
\hline
\end{tabular}
\end{center}
\end{table}
%%%%%%%%%%%%%%%%%%%%%%%%%%%%%%%%%%%%%%

Note that in our electronic structure calculations, we rely on experimentally
measured atomic positions and lattice constants, because they are well
established for this material and still more accurate than those
obtained from DFT.

\subsection{Calculation details}

The details of the computational method are described in our previous papers
\cite{AJY+06,AHY+07b,AYJ10,AKB22a} and here we only mention several
aspects. The band structure calculations were performed using the fully
relativistic LMTO method \cite{And75,book:AHY04}. This implementation of the
LMTO method uses four-component basis functions constructed by solving the
Dirac equation inside an atomic sphere \cite{NKA+83}. The exchange-correlation
functional of a GGA-type was used in the version of Perdew, Burke and
Ernzerhof \cite{PBE96}. The Brillouin zone integration was performed using the
improved tetrahedron method \cite{BJA94}. The basis consisted of Ir and Ba
$s$, $p$, $d$, and $f$; and In and O $s$, $p$, and $d$ LMTO's.

To consider the electron-electron correlation effects we used the
relativistic generalization of the rotationally invariant version of the
GGA+$U$ method \cite{YAF03} which considers that, in the presence of
spin-orbit coupling, the occupation matrix of localized electrons becomes
nondiagonal in spin indexes. Hubbard $U$ was considered an external parameter
and varied from 0.65 to 3.65 eV.  We used in our calculations the value of
exchange Hund coupling $J_H$=0.65 eV obtained from constrained LSDA
calculations \cite{DBZ+84,PEE98}. Thus, the parameter $U_{\rm{eff}}=U-J_H$,
which roughly determines the splitting between the lower and upper Hubbard
bands, varied between 0 and 3.0 eV. We adjusted the value of $U$ to achieve
the best agreement with the experiment.

In the RIXS process, an electron is promoted from a core level to an
intermediate state, leaving a core hole. As a result, the electronic structure
of this state differs from that of the ground state. To reproduce the
experimental spectrum, the self-consistent calculations should be carried out
including a core hole. Usually, the core-hole effect has no impact on the
shape of XAS at the $L_{2,3}$ edges of 5$d$ systems and just a minor effect on
the XMCD spectra at these edges \cite{book:AHY04}. However, the core hole has
a strong effect on the RIXS spectra in transition metal compounds
\cite{AKB22a,AKB22b}; therefore, we consider it in our calculations.

\section{Electronic and magnetic structures}
\label{sec:bands}

%%%%%%%%%%%%%%%%%%%%%%%%%%%%
\begin{figure}[tbp!]
\begin{center}
\includegraphics[width=0.99\columnwidth]{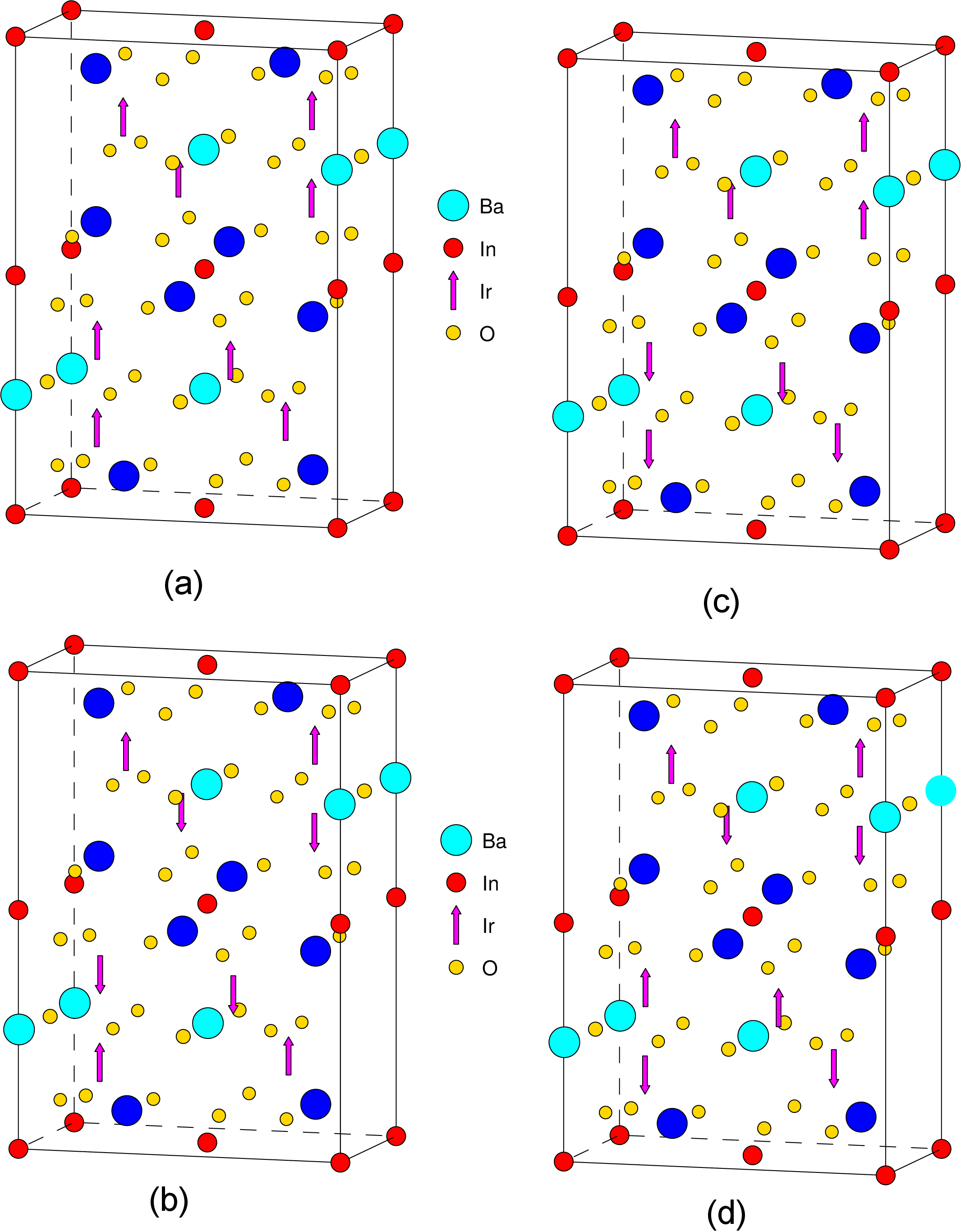}
\end{center}
\caption{\label{magn_str_BIIO}(Color online) Different magnetic
  ordering in Ba$_3$InIr$_2$O$_9$ with the $C2/c$ crystal structure:
  ferromagnetic (a), antifferromagnetic AFM$_1$ (b), AFM$_2$ (c), and
  AFM$_3$ (d). }
\end{figure}
%%%%%%%%%%%%%%%%%%%%%%%%

We performed GGA, GGA+SO, and GGA+SO+$U$ calculations of the
electronic and magnetic structures of Ba$_3$InIr$_2$O$_9$ for the
experimental crystal structure \cite{DMO+17} (see Table
\ref{struc_tab_BIIO}). In order to understand the magnetic properties
of the system, four different magnetic configurations, namely FM (both
the intra- and inter-dimer couplings are FM), AFM$_1$ (both the intra-
and inter-dimer couplings are AFM), AFM$_2$ (the intra-dimer coupling
is FM and the inter-dimer coupling is AFM), and AFM$_3$ (the
intra-dimer coupling is AFM and the inter-dimer coupling is FM), have
been simulated (see Fig. \ref{magn_str_BIIO}). The calculations show
that the AFM$_2$ state has the lowest energy (Table
\ref{Etot_BIIO}). A similar magnetic ground state was obtained by
Panda {\it et al.} for Ba$_2$YIr$_2$O$_9$ \cite{PBL+15} and Nag {\it
  et al.} for Ba$_2$ZnIr$_2$O$_9$ \cite{NMB+16}.

The difference in total energy between magnetic and nonmagnetic solutions
(with considering SOC and Hubbard correlations) is relatively large and equals
to 1.682 eV per formula unit. However, between different magnetic solutions 
it is very small. Probably, it may explain the fact that Ba$_3$InIr$_2$O$_9$ 
possesses no long-range magnetic order down to very low temperatures and 
points out on possible spin-liquid behavior of local moments below 1 K 
in this perovskite \cite{DMO+17}. We also found several noncollinear 
AFM solutions but with significantly higher total energy than the solutions 
with FM and AFM ordering along the $c$ direction.

%%%%%%%%%%%%%%%%%%%%%%%%%%%%%%%%%%%%%%%%%%%%%%%%%%%%%%%
\begin{table}[tbp!]
  \caption{\label{Etot_BIIO} The total energy $E_{total}$ per atom (in meV) in
    Ba$_3$InIr$_2$O$_9$ with the $C2/c$ crystal structure calculated in the
    GGA+SO+$U$ approach for the FM, AFM$_1$, AFM$_2$, and AFM$_3$
    configurations. }
\begin{center}
\begin{tabular}{ccccccc}
\hline
 FM  & AFM$_1$ & AFM$_2$ & AFM$_3$  \\
\hline
 0.232 & 3.375 & 0 & 4.273  \\
\hline
\end{tabular}
\end{center}
\end{table}
%%%%%%%%%%%%%%%%%%%%%%%%%%%%%%%%%%%%%%%%%%%%%%%%%%%%%%%%

%%%%%%%%%%%%%%%%%%%%%%%%%%%%
\begin{figure}[tbp!]
\begin{center}
\includegraphics[width=0.99\columnwidth]{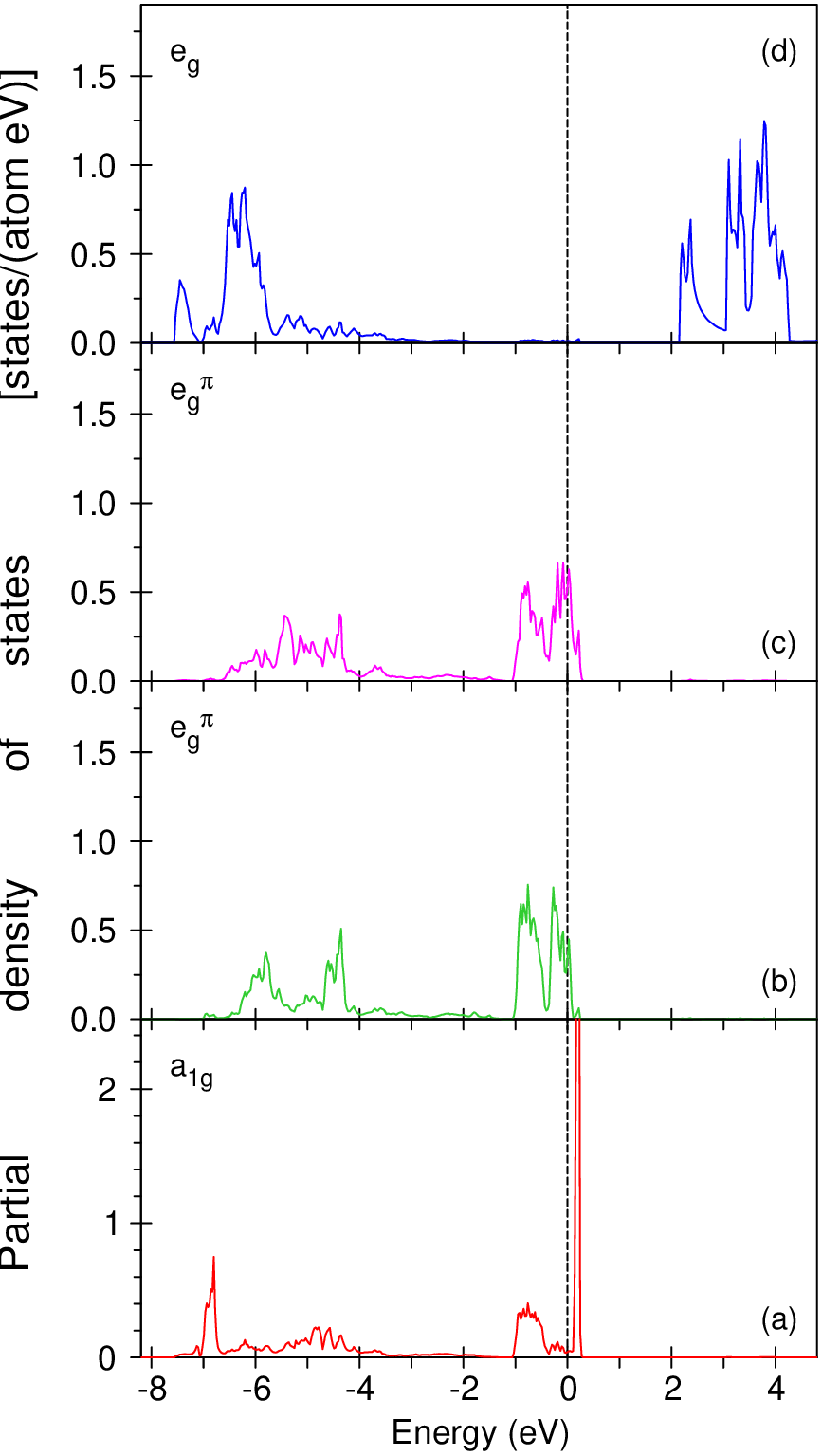}
\end{center}
\caption{\label{Orbitals_BIIO}(Color online) The orbital resolved
  partial density of states (DOS) [in states/(atom eV)] for
  Ba$_3$InIr$_2$O$_9$ calculated in the GGA. }
\end{figure}
%%%%%%%%%%%%%%%%%%%%%%%%

In Figure \ref{Orbitals_BIIO} the plots of orbital resolved partial
density of states (DOS) for Ba$_3$InIr$_2$O$_9$ are presented. For the
dimers formed by two face-shared IrO$_6$ octahedra (see
Fig. \ref{struc_BIIO}), the local distortion of $D_{3d}$ symmetry
leads to the splitting of {\tg} orbitals into an $a_{1g}$ singlet and
an $e_g^{\pi}$ doublet. The original {\eg} ($e_g^{\sigma}$ ) doublet
remains unsplit. The largest bonding-antibonding splitting corresponds
to $a_{1g}$ orbitals, directed to each other in this geometry. The
rightmost peak in Fig. \ref{Orbitals_BIIO}(a) corresponds to the
antibonding $a_{1g}$ orbital at 0.2 eV, while the antibonding
$e_g^{\pi}$ states are centered at $\sim$ -0.4 to 0.1 eV
[Fig. \ref{Orbitals_BIIO}(b, c)]. Due to the strong distortion of
IrO$_6$ octahedra the {\tg} orbitals are strongly mixed. In the local
coordinate system with $z$ directed from the Ir ion to the O$_5$ one
[see Fig. \ref{struc_BIIO}(b)], the $a_{1g}$ molecular orbital can be
presented as $a_{1g}$ = 0.523{\dxy}+0.275{\dyz}+0.712{\dxz}. There is
competition between the SO entanglement and the dimerization in
Ba$_3$InIr$_2$O$_9$. The former mixes up different orbital states to
create the orbital moment. The latter selects a specific orbital to
gain bonding energy. We note that the effect of electron correlations
may narrow the bandwidth and make the RIXS peaks sharper.

The dimerization of transition-metal ions has been frequently seen, for
example, in high pressure honeycomb-based 5$d^5$ iridates
\cite{AUU18,TKG+19,AKU+21} and in a wide variety of honeycomb-based 3$d$ and
4$d$ oxides and halides, including $\alpha$-TiCl$_3$ (3$d^1$) \cite{Oga60},
$\alpha$-MoCl$_3$ (4$d^3$) \cite{MYL+17}, Li$_2$RuO$_3$ (4$d^4$)
\cite{MYS+07}, and $\alpha$-RuCl$_3$ \cite{AKK+17}.

%%%%%%%%%%%%%%%%%%%%%%%%%%%%
\begin{figure}[tbp!]
\begin{center}
\includegraphics[width=0.99\columnwidth]{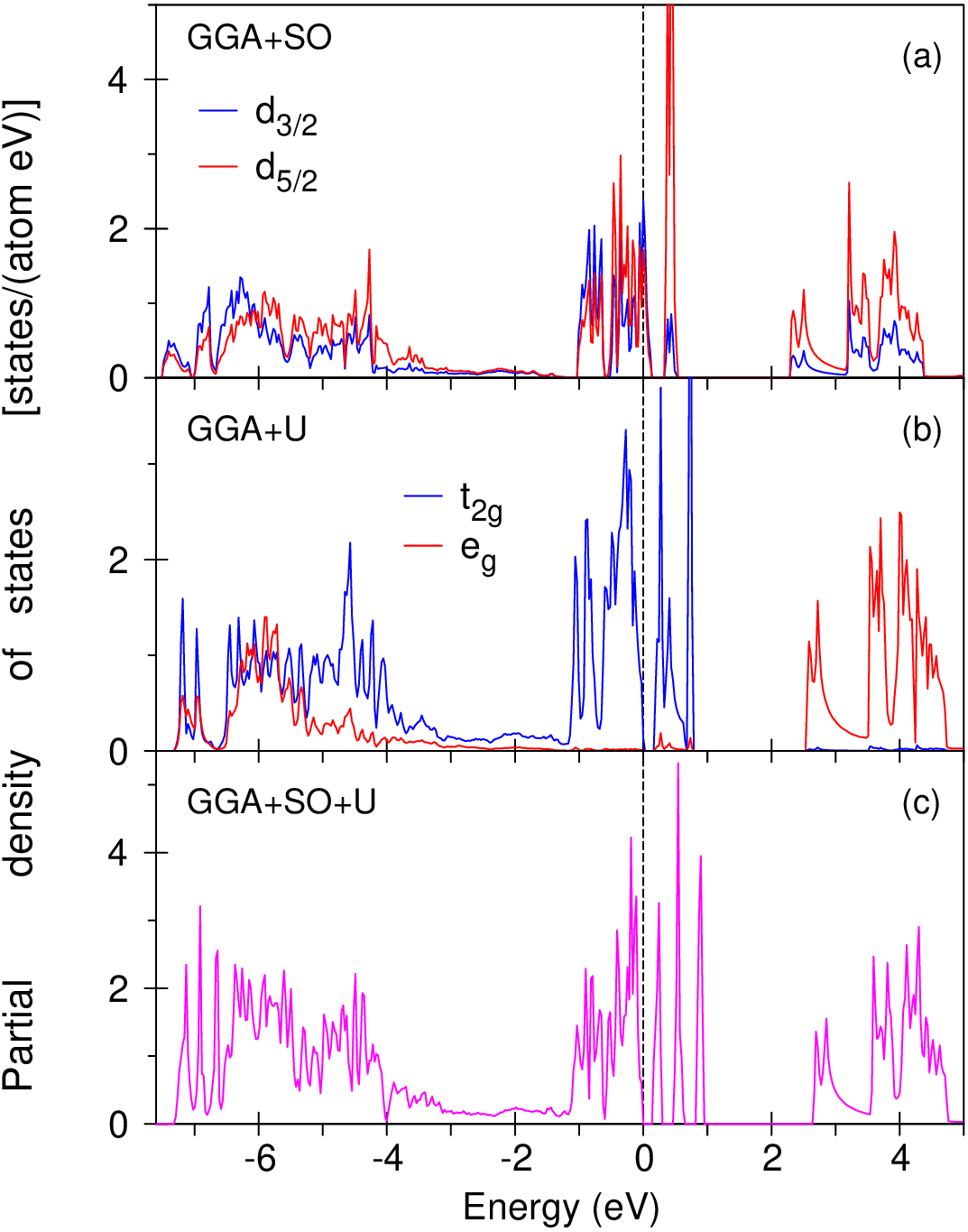}
\end{center}
\caption{\label{PDOS_BIIO_SO_U}(Color online) The Ir 5$d$ partial
  density of states (DOS) [in states/(atom eV)] for
  Ba$_3$InIr$_2$O$_9$ calculated in the GGA+SO (a), GGA+$U$ (b), and
  GGA+SO+$U$ ($U_{\rm{eff}}$= 2.3 eV) (c) approaches. }
\end{figure}
%%%%%%%%%%%%%%%%%%%%%%%%

According to the experimental measurements of Revelli {\it et al.}
\cite{RSM+22}, the real part of the conductivity is $\sigma \sim$
10$^{-6}$/$\Omega$ cm, which firmly indicates the insulating character
of Ba$_3$InIr$_2$O$_9$. This conclusion is also supported by the
observation that the dielectric loss $\varepsilon'$ is much smaller
than the permittivity $\varepsilon''$, as usually expected for an
insulator \cite{RSM+22}.  Figure \ref{PDOS_BIIO_SO_U} shows the Ir
5$d$ partial DOS for Ba$_3$InIr$_2$O$_9$ calculated in the GGA+SO,
GGA+$U$, and GGA+SO+$U$ approaches. The GGA and GGA+SO approaches give
a metallic ground state in Ba$_3$InIr$_2$O$_9$. It is in contradiction
with the experiment. To produce an insulating ground state we used the
GGA+SO+$U$ method. We discovered that the energy gap opens up for
$U_{\rm{eff}}$ = 1.5 eV for the AFM ground state. The calculations
show that the energy gap is already formed in the GGA+$U$
approximation without SOC. Ba$_3$InIr$_2$O$_9$ can be classified as a
Mott insulator, since it was expected to be metallic from GGA band
structure calculations. We found that the best agreement between the
calculated and experimentally measured RIXS spectra at the Ir $L_3$
edge can be achieved for $U_{\rm{eff}}$ = 2.3 eV (see Section IV).

\begin{figure}[tbp!]
\begin{center}
\includegraphics[width=0.99\columnwidth]{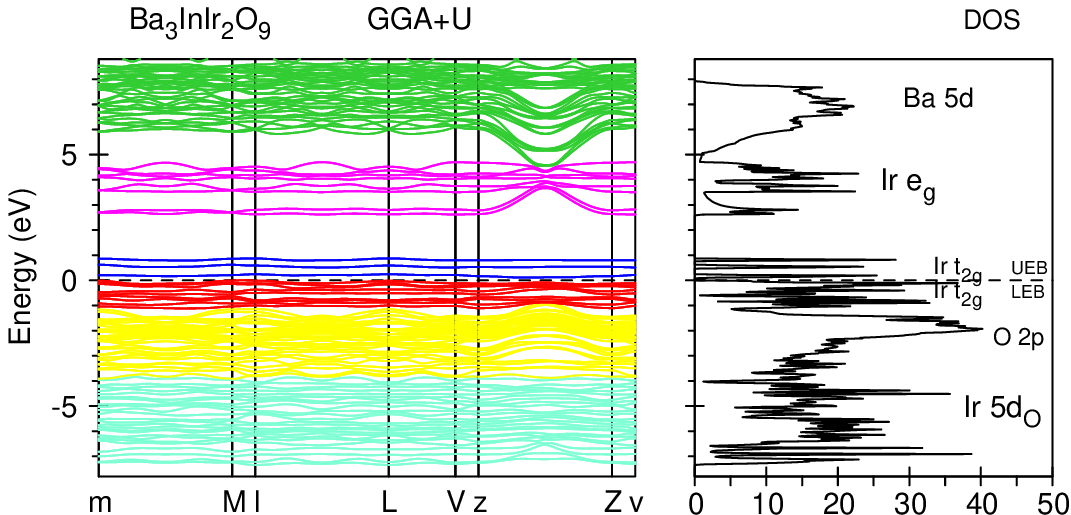}
\end{center}
\caption{\label{BND_BIIO}(Color online) The energy band structure and
  total density of states (DOS) [in states/(cell eV)] for
  Ba$_3$InIr$_2$O$_9$ calculated in the GGA+SO+$U$ approach
  ($U_{\rm{eff}}$= 2.3 eV). }
\end{figure}

\begin{figure}[tbp!]
\begin{center}
\includegraphics[width=0.9\columnwidth]{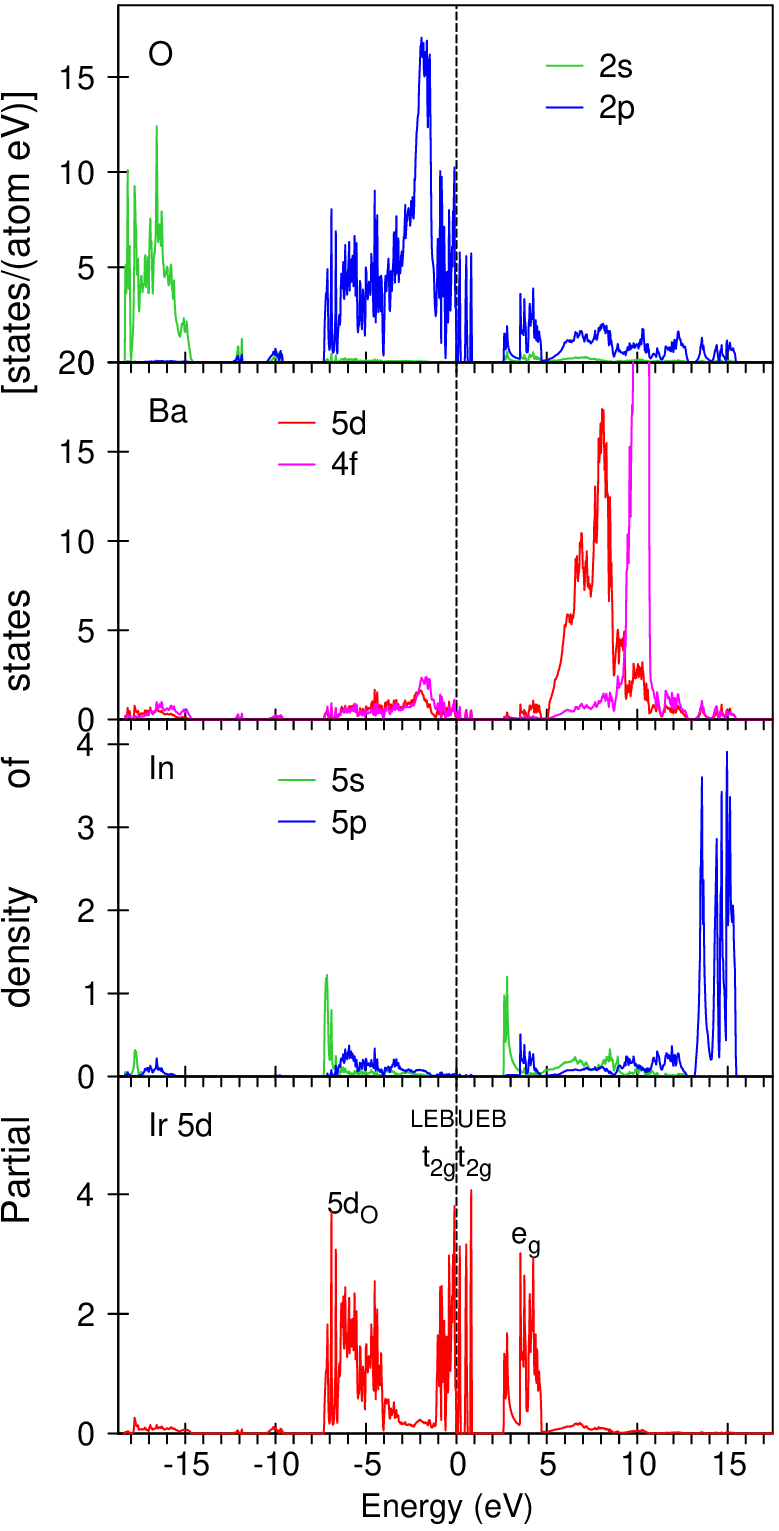}
\end{center}
\caption{\label{PDOS_BIIO}(Color online) The partial density of states
  (DOS) [in states/(atom eV)] for Ba$_3$InIr$_2$O$_9$ calculated in
  the GGA+SO+$U$ approach ($U_{\rm{eff}}$ = 2.3 eV). }
\end{figure}

Figures \ref{BND_BIIO} and \ref{PDOS_BIIO} present the energy band
structure and partial DOSs, respectively, in Ba$_3$InIr$_2$O$_9$ for
the AFM$_2$ solution calculated in the GGA+SO+$U$ approach with
$U_{\rm{eff}}$ = 2.3 eV. The occupied {\tg} states [the low energy
  band (LEB)] are situated in the energy interval from $-$1.2 eV to
$E_F$ and can be subdivided into two band groups from $-$1.2 to $-$0.6
eV and from $-$0.6 eV to $E_F$. The empty {\tg} states [the upper
  energy band (UEB)] consist of three narrow single peaks divided by
energy gaps and occupy the energy range from 0.07 to 0.88 eV. There is
a significant amount of Ir 5$d$ DOS located at the bottom of oxygen
2$p$ states from $-$7.35 to $-$4.05 eV below the Fermi energy. These,
so called, Ir 5$d_{\rm{O}}$ states are provided by the tails of oxygen
2$p$ states inside the Ir atomic spheres and play an essential role in
the RIXS spectrum at the Ir $L_3$ edge (see Section IV).

The In 5$p$ states are mostly situated above the Fermi level from 13.2
to 15.5 eV and consist of four separated energy peaks. The Ba 5$d$
states occupy the energy region from 5.1 to 9.2 eV above the energy
Fermi. A narrow and intensive DOS peak of Ba 4$f$ states is located
just above the Ba 5$d$ states from 9.2 to 10.8 eV. The oxygen 2$s$
states are situated far below the Fermi level from $-$18.3 to $-$14.6
eV. The occupied O 2$p$ states are localized from $-$7.35 eV to
$E_F$. They are strongly hybridized with Ir 5$d$ states from $-$7.35
eV to $-$4.05 eV and from $-$1.2 eV to $E_F$. The empty oxygen 2$p$
states are strongly hybridized with Ir {\tg} UEB just above the Fermi
level and with the Ir {\eg} states from 2.6 to 4.7 eV. They are also
hybridized with Ba 5$d$ and 4$f$ states as well as In 5$p$ states.

%%%%%%%%%%%%%%%%%%%%%%%%%%%%
%%%%%%%%%%%%%%%%%%%%%%%%

%%%%%%%%%%%%%%%%%%%%%%%%%%%%
%%%%%%%%%%%%%%%%%%%%%%%%

%%%%%%%%%%%%%%%%%%%%%%%%%%%%%%%%%%%%%%%%%%%%%%%%%%%%%%%
\begin{table}[tbp!]
  \caption{\label{mom_BIIO} The theoretically calculated in
    the GGA+SO+$U$ approach ($U_{\rm{eff}}$ = 2.3 eV) spin $M_s$,
    orbital $M_l$, and total $M_{tot}$ magnetic moments (in {\mb}) in
    Ba$_3$InIr$_2$O$_9$ for the AFM$_{001}$ solution. }
\begin{center}
\begin{tabular}{ccccccc}
\hline
 atom  & $M_s$ & $M_l$ &  $M_{tot}$ \\
\hline
 Ba$_1$ & 0.0415 & $-$0.0020 & 0.0395  \\
 Ba$_2$ & 0.0141 & 0.0056    & 0.0197  \\
 In     & 0 & 0 & 0  \\
 Ir     & 0.3903 & 0.1393 & 0.5296  \\
 O$_1$  & 0.1188 & 0.0397 & 0.1585  \\
 O$_2$  & 0.0386 & 0.0054 & 0.0440  \\
 O$_3$  & 0.0513 & 0.0133 & 0.0646  \\
 O$_4$  & 0.0430 & 0.0015 & 0.0445  \\
 O$_5$  & 0.0667 & 0.0131 & 0.0798  \\
\hline
\end{tabular}
\end{center}
\end{table}
%%%%%%%%%%%%%%%%%%%%%%%%%%%%%%%%%%%%%%%%%%%%%%%%%%%%%%%%

Table \ref{mom_BIIO} presents the theoretically calculated in the GGA+SO+$U$ 
approach ($U_{\rm{eff}}$ = 3.2 eV) spin $M_s$, orbital $M_l$, and total $M_{tot}$ 
magnetic moments in Ba$_3$InIr$_2$O$_9$ for the AFM$_{001}$ solution. 
The spin and orbital moments at the Ir site in Ba$_2$InIrYO$_9$
equal to 0.3903 and 0.1393 {\mb}, respectively. They are in the same direction, 
as Ir 5$d$ states are more than half filled. The ratio $M_l/M_s\sim$0.357 is relatively small 
compared to other SOC driven $J_{\rm{eff}}$ = $\frac{1}{2}$ iridates, i.e., Sr$_2$IrO$_4$,
where it is $\sim$1.68 \cite{AKB24}. It may indicate only a moderate SOC
influence in Ba$_2$InIr$_2$O$_9$.

\section{I\lowercase{r} RIXS spectra}
\label{sec:rixs}

\begin{figure}[tbp!]
\begin{center}
\includegraphics[width=0.9\columnwidth]{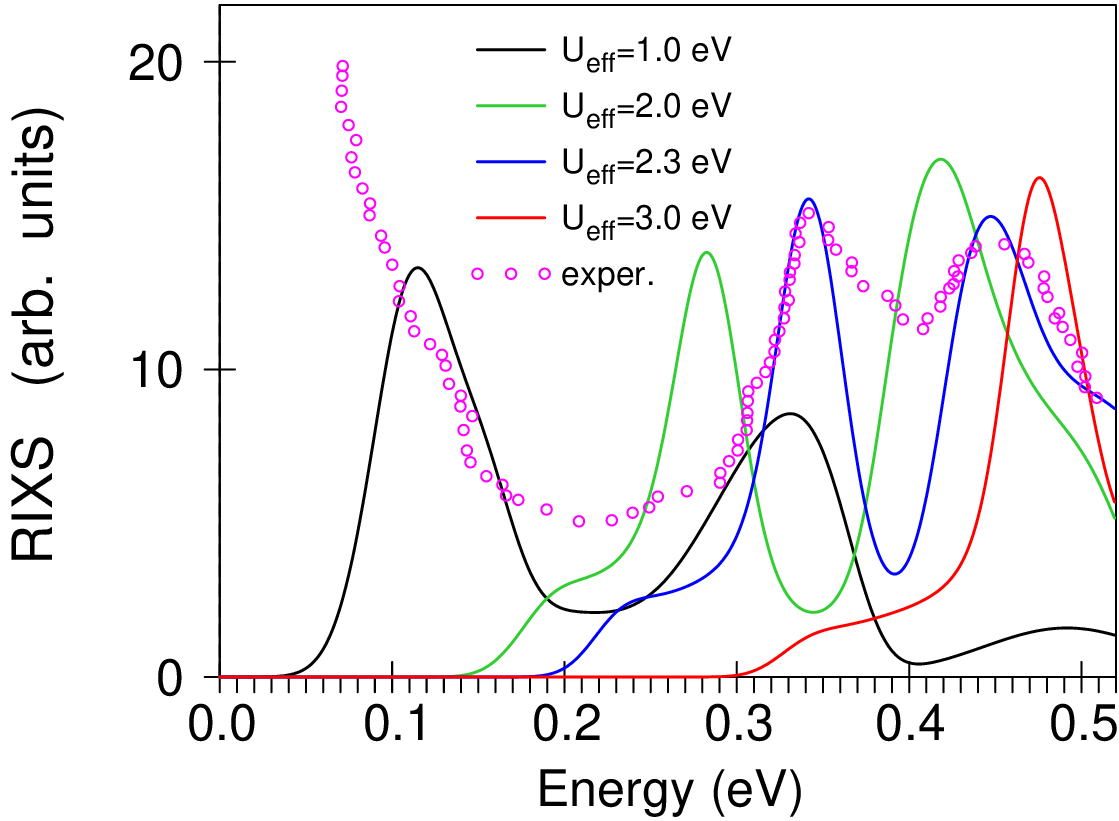}
\end{center}
\caption{\label{rixs_U_BIIO}(Color online) The experimental resonance
  inelastic x-ray scattering (RIXS) spectrum (open magenta circles)
  measured by Revelli {\it et al.}  \cite{RSM+22} at the Ir $L_3$ edge
  in Ba$_3$InIr$_2$O$_9$ compared with the theoretically calculated
  one in the GGA+SO+$U$ approach for different $U_{\rm{eff}}$. }
\end{figure}

The experimental RIXS spectrum at the Ir $L_3$ edge was measured by
Revelli {\it et al.}  \cite{RSM+22} in the energy range up to 5.5
eV. In addition to the elastic peak centered at zero energy loss the
spectrum consists of several peaks below 2 eV, a strong peak at 3.9
eV, and a fine structure above 4.4 eV. We found that the fine
structure situated below 2 eV corresponds to intra-{\tg}
excitations. These peaks are very sensitive to the value of the energy
gap in Ba$_3$InIr$_2$O$_9$ and the relative position of {\tg} LEB and
UEB (Fig. \ref{BND_BIIO}). Figure \ref{rixs_U_BIIO} shows the
experimental RIXS spectrum measured by Revelli {\it et al.}
\cite{RSM+22} compared with the theoretical spectra calculated for
$\tg \to \tg$ transitions in the GGA+SO+$U$ approach for the
AFM$_{001}$ solution for different $U_{\rm{eff}}$ values. The best
agreement was found for $U_{\rm{eff}}$ = 2.3 eV. The GGA+SO
calculations (not shown) as well as the GGA+SO+$U$ approach with
smaller $U_{\rm{eff}}$ do not produce adequate agreement with the
experimental data.  The larger values of $U_{\rm{eff}}$ shift the RIXS
spectrum towards higher energies.

%%%%%%%%%%%%%%%%%%%%%%%%%%%%
%%%%%%%%%%%%%%%%%%%%%%%%

\begin{figure}[tbp!]
\begin{center}
\includegraphics[width=0.9\columnwidth]{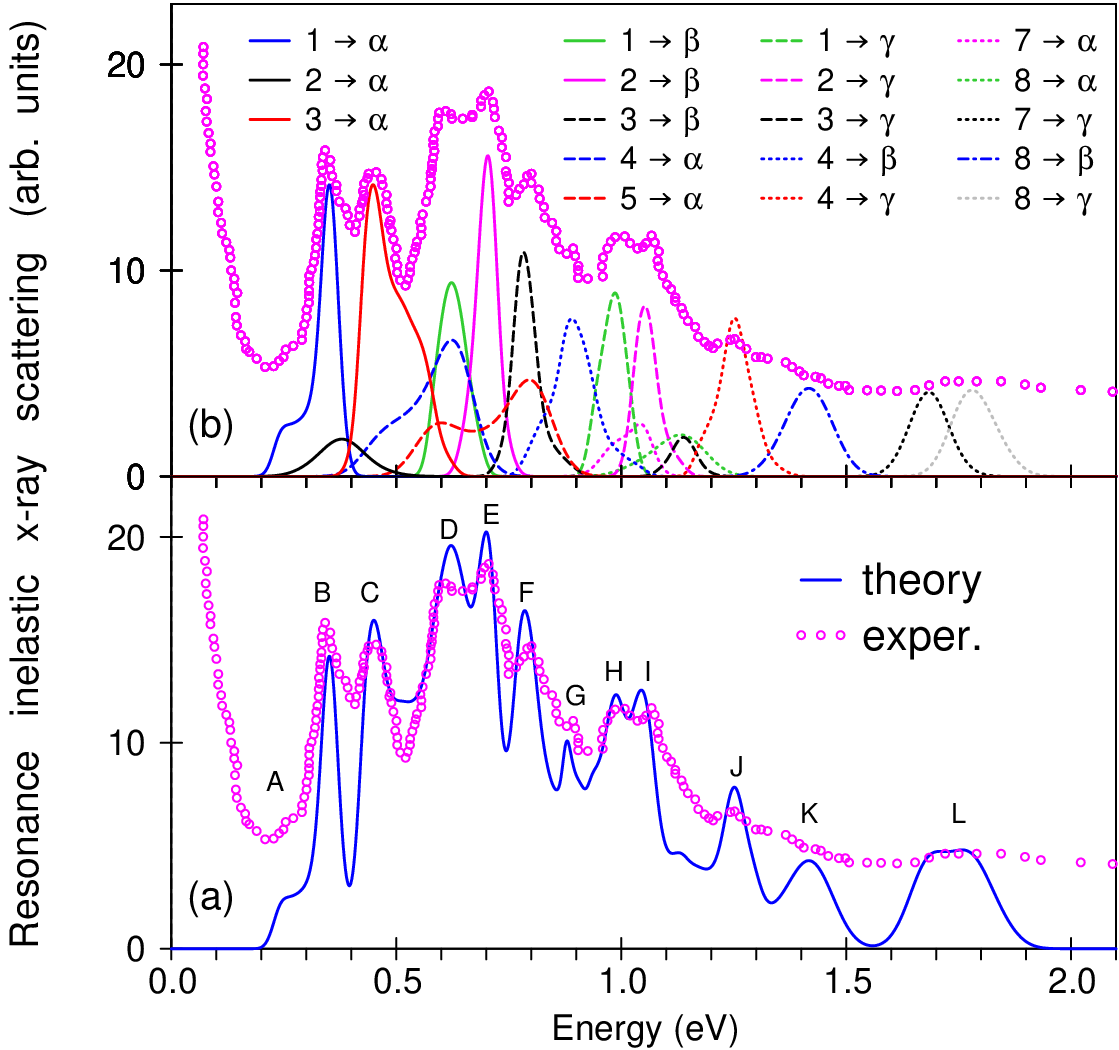}
\end{center}
\caption{\label{rixs_t2g_BIIO}(Color online) The experimental
  resonance inelastic x-ray scattering (RIXS) spectrum (open magenta
  circles) measured by Revelli {\it et al.}  \cite{RSM+22} at the Ir
  $L_3$ edge in Ba$_3$InIr$_2$O$_9$ compared with (a) the
  theoretically calculated one in the GGA+SO+$U$ approach
  ($U_{\rm{eff}}$ = 2.3 eV) for {\tg} $\rightarrow$ {\tg} transitions;
  (b) partial contributions from different transitions between the
  bands presented in Fig. \ref{BND_t2g_BIIO}. }
\end{figure}

\begin{figure}[tbp!]
\begin{center}
\includegraphics[width=1.\columnwidth]{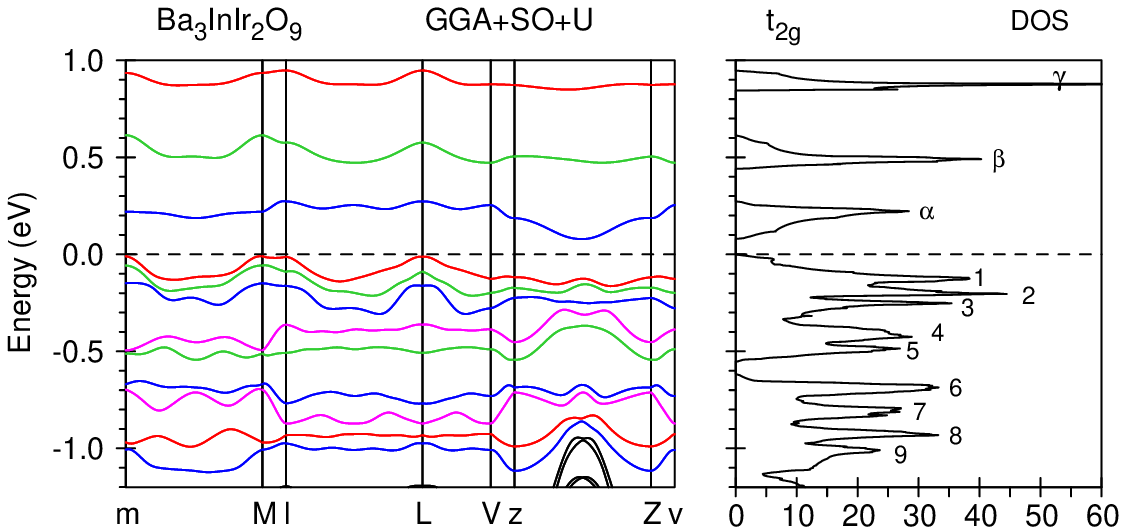}
\end{center}
\caption{\label{BND_t2g_BIIO}(Color online) The energy band structure 
  of Ir {\tg} states calculated for Ba$_3$InIr$_2$O$_9$ in the GGA+SO+$U$ 
  approach ($U_{\rm{eff}}$=2.3 eV). }
\end{figure}

Figure \ref{rixs_t2g_BIIO}(a) shows the theoretical RIXS spectrum at
the Ir $L_3$ edge for the {\tg} $\rightarrow$ {\tg} transitions in
Ba$_3$InIrO$_9$ compared with the measurements of Revelli {\it et al.}
\cite{RSM+22}. The Ir $L_3$ spectrum for the {\tg} $\rightarrow$ {\tg}
transitions has quite a rich fine structure with at least twelve well
separated peaks from $A$ to $L$ below 2 eV.  The appearance of
multiple peaks in the RIXS spectrum is a direct consequence of strong
noncubic crystal field (CF) splitting originating from the distorted
octahedral environment of the Ir ions. The CF splitting leads to a
very specific energy band structure of Ir {\tg} states presented in
Fig. \ref{BND_t2g_BIIO}. There are nine {\tg} bands below $E_F$ (from
1 to 9) and three empty bands $\alpha$, $\beta$, and $\gamma$
separated by energy gaps. The interband transitions between these nine
occupied and three empty bands produce 27 peaks. Some of these peaks
are quite small. As a result, the structure of the Ir $L_3$ RIXS
spectrum below 2 eV consists of at least twelve well distinguished
peaks. The corresponding interband transitions are presented in
Fig. \ref{rixs_t2g_BIIO}(b). The low energy peaks $A$, $B$, and $C$
are due to the transitions from the first, second, and third occupied
bands (the red, green, and blue bands in Fig. \ref{BND_t2g_BIIO}) into
the low energy empty band (the blue band with DOS peak $\alpha$). The
other peaks from $D$ to $L$ are due to many interband transitions,
which are presented in Fig. \ref{rixs_t2g_BIIO}(b).

%%%%%%%%%%%%%%%%%%%%%%%%%%%%
%%%%%%%%%%%%%%%%%%%%%%%%

%%%%%%%%%%%%%%%%%%%%%%%%%%%%
%%%%%%%%%%%%%%%%%%%%%%%%

\begin{figure}[tbp!]
\begin{center}
\includegraphics[width=0.9\columnwidth]{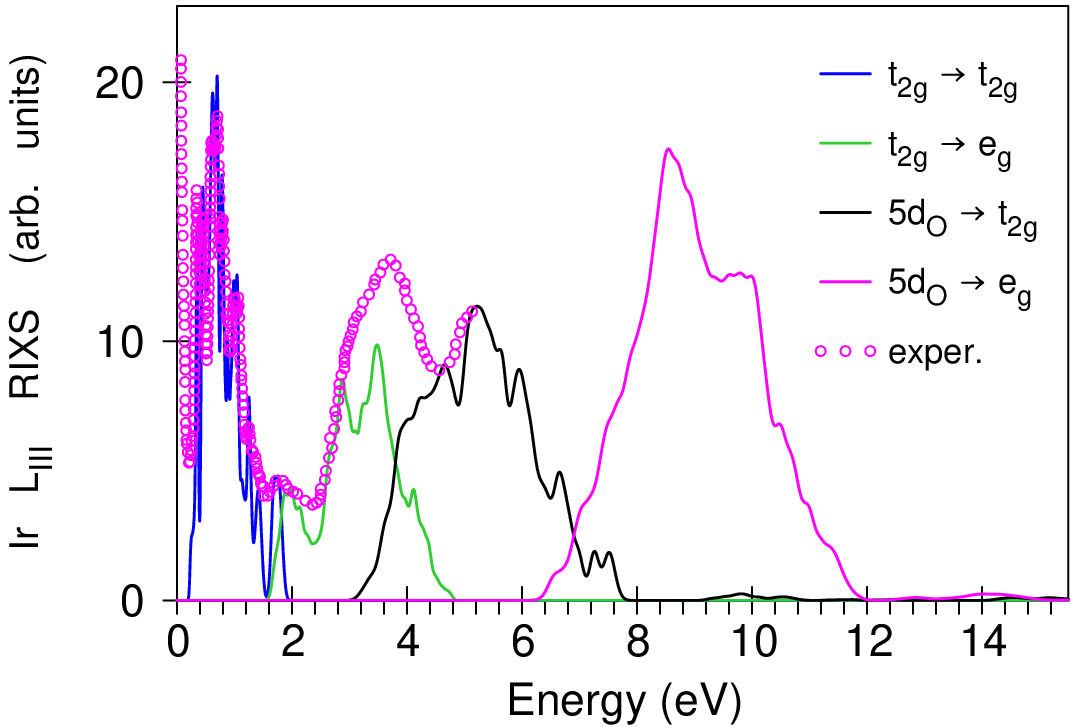}
\end{center}
\caption{\label{rixs_Ir_L3_BIIO}(Color online) The experimental
  resonance inelastic x-ray scattering (RIXS) spectrum (open magenta
  circles) measured by Revelli {\it et al.}  \cite{RSM+22} at the Ir
  $L_3$ edge in Ba$_3$InIr$_2$O$_9$ compared with the theoretically
  calculated one in the GGA+SO+$U$ approach ($U_{\rm{eff}}$=2.3 eV) in
  a wide energy range. }
\end{figure}

Figure \ref{rixs_Ir_L3_BIIO} presents the theoretically calculated and
experimentally measured RIXS spectra at the Ir $L_3$ edge for
Ba$_3$InIr$_2$O$_9$ \cite{RSM+22} in a wide energy interval up to 14
eV. As we mentioned above, the peaks situated below 2 eV correspond to
intra-{\tg} excitations (the blue curve in
Fig. \ref{rixs_Ir_L3_BIIO}). The peak located at $\sim$4 eV was found
to be due to $\tg \rightarrow \eg$ transitions (the green curve). The
increase of the intensity above 4.5 eV can be associated with
charge-transfer excitations 5$d_{\rm{O}}$ $\rightarrow \tg$ (the black
curve). These transitions also contribute into the major peak at 4
eV. The strong fine structure at $\sim$9 eV (the magenta curve) is due
to 5$d_{\rm{O}}$ $\rightarrow$ {\eg} transitions. The theoretical
calculations are in good agreement with the experimental data.

%%%%%%%%%%%%%%%%%%%%%%%%%%%%
%%%%%%%%%%%%%%%%%%%%%%%%

%%%%%%%%%%%%%%%%%%%%%%%%%%%%
\begin{figure}[tbp!]
\begin{center}
\includegraphics[width=0.9\columnwidth]{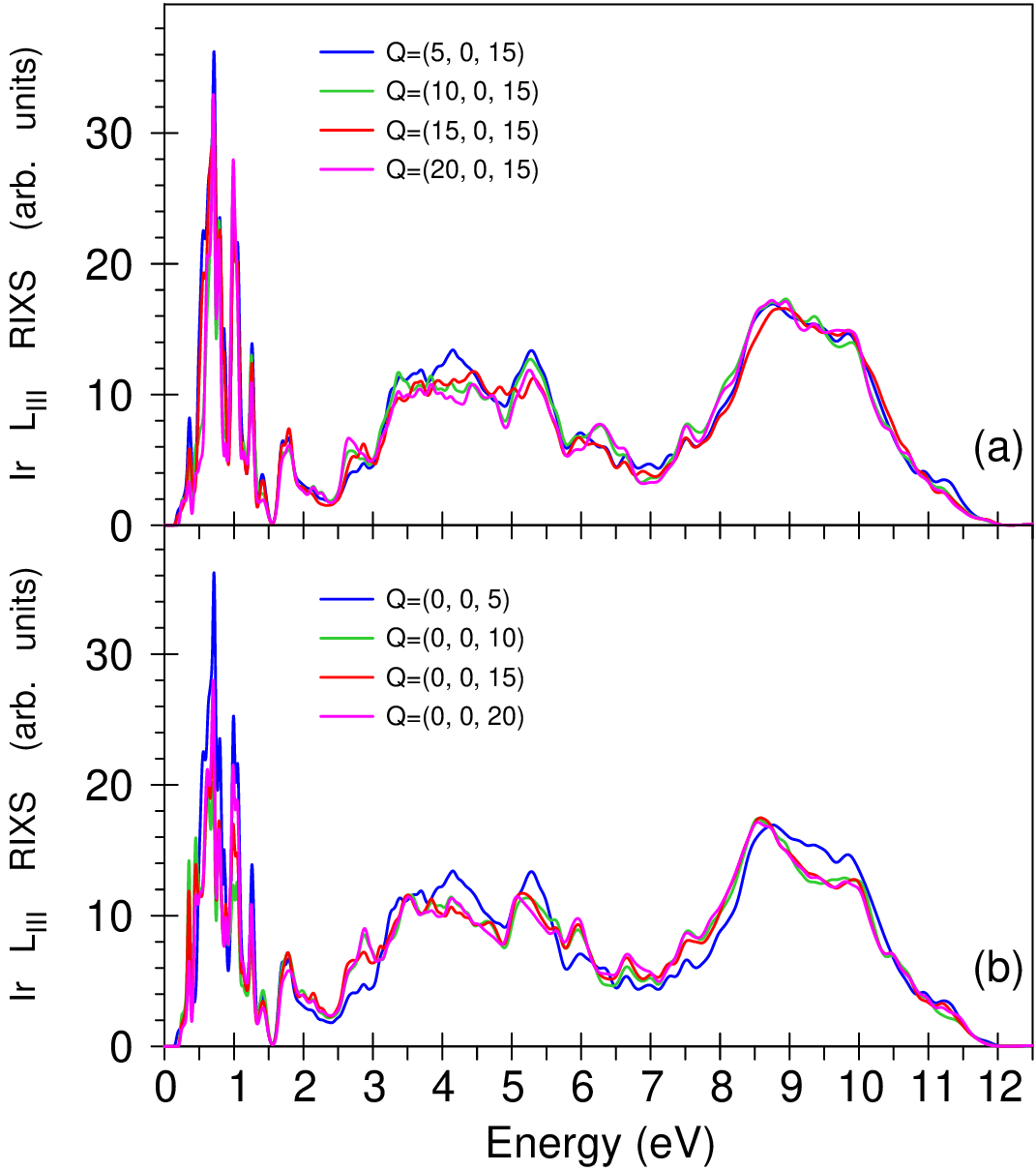}
\end{center}
\caption{\label{rixs_Ir_Qxz_BIIO}(Color online) The resonance
  inelastic x-ray scattering (RIXS) spectra at the Ir $L_3$ edge in
  Ba$_3$InIr$_2$O$_9$ calculated as a function of $Q_x$ with the
  momentum transfer vectors {\bf Q} = (Q$_x$, 0, 15) in reciprocal
  lattice units (the upper panel) and $Q_z$ with {\bf Q} = (0, 0,
  Q$_z$) in reciprocal lattice units (the lower panel) for incident
  photon energy $\hbar \omega_{in}$ = 11215 eV. }
\end{figure}
%%%%%%%%%%%%%%%%%%%%%%%%

It is widely believed that $d-d$ excitations show only small momentum
transfer vector {\bf Q} dependence in 5$d$ transition metal compounds
\cite{LKH+12,KTD+20}. As we see in the upper panel of
Fig. \ref{rixs_Ir_Qxz_BIIO}, the RIXS spectra show small dependence as
a function of $Q_x$ with the momentum transfer vector {\bf Q} =
(Q$_x$, 0, 15) in reciprocal lattice units (r.l.u.) for incident
photon energy $\hbar \omega_{in}$ = 11215 eV in the energy interval up
to 12 eV. The lower panel of Fig. \ref{rixs_Ir_Qxz_BIIO} shows the
RIXS spectra at the Ir $L_3$ edge in Ba$_3$InIr$_2$O$_9$ calculated as
a function of $Q_z$ with the momentum transfer vector {\bf Q} = (0, 0,
Q$_z$) r.l.u.  We found that the RIXS spectrum possesses an
oscillation character. With increasing Q$_z$ from 5 to 10, 15, and 20
r.l.u. the low energy structure, which reflects the $\tg \rightarrow
\tg$ transitions, is increased then decreased, again increased and
after that decreased. The high energy fine structures above 3 eV are
changed insignificantly with the increase of Q$_z$. Similar dependence
of the RIXS spectrum on Q$_z$ was detected experimentally by Revelli
{\it et al.} \cite{RSM+22}. Namely, they found that the low energy
RIXS spectrum below 1.5 eV show an oscillation character for the
momentum transfer vector {\bf Q} = (0.3, 0, Q$_z$). It is increased
with the changing of Q$_z$ from 11.2 to 14.0, then decreased for Q$_z$
= 16.8, and increased for Q$_z$ = 19.6 r.l.u. Although, the peaks
above 3 eV were found to be almost insensible to the changing of
Q$_z$. Analyzing Fig. \ref{rixs_Ir_Qxz_BIIO}, we can conclude that the
momentum dependence of the excitations in Ba$_3$InIr$_2$O$_9$ is
rather small, as it was earlier observed in other iridates, such as
Sr$_3$CuIrO$_6$ \cite{LKH+12}, In$_2$Ir$_2$O$_7$ \cite{KTD+20} or
Sr$_2$IrO$_4$ \cite{AKB24}.

%%%%%%%%%%%%%%%%%%%%%%%%%%%%
\begin{figure}[tbp!]
\begin{center}
\includegraphics[width=0.9\columnwidth]{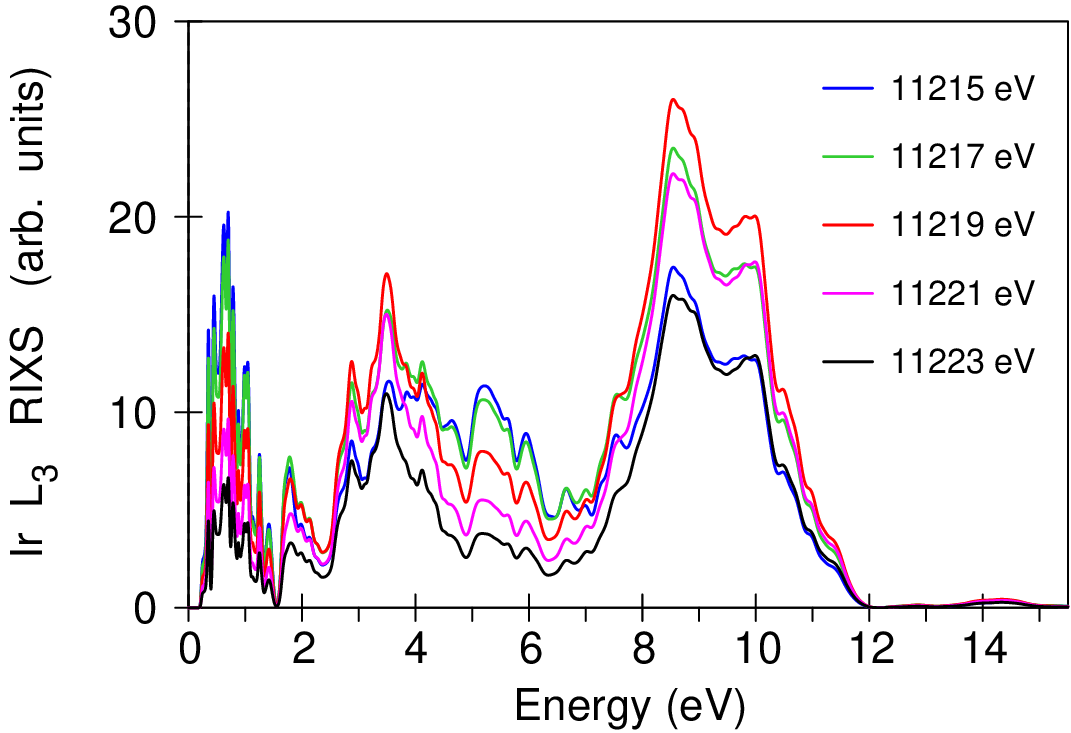}
\end{center}
\caption{\label{rixs_Ei_BIIO}(Color online) The resonance inelastic
  x-ray scattering (RIXS) spectra as a function of incident photon
  energy $E_i$ calculated at the Ir $L_3$ edge in Ba$_3$InIr$_2$O$_9$
  with the momentum transfer vector {\bf Q} = (0, 0, 15) in reciprocal
  lattice units. }
\end{figure}
%%%%%%%%%%%%%%%%%%%%%%%%

Figure \ref{rixs_Ei_BIIO} shows the Ir $L_3$ RIXS spectrum as a function of
incident photon energy $E_i$ above the corresponding edge with the momentum
transfer vector {\bf Q} = (0, 0, 15). We found that the low energy fine
structure corresponding to the intra-{\tg} excitations steadily decreases when
the incident photon energy changes from 11215 to 11223 eV, whereas the energy
peak corresponding to the $\tg \rightarrow \eg$ transitions firstly increases,
when the incident photon energy changes from 11215 to 11219 eV, and after that 
decreases. The same behavior occurs for the high energy peak between 7 and 
12 eV, which corresponds to the charge transfer transitions 
5$d_{\rm{O}}$ $\rightarrow \eg$.

%%%%%%%%%%%%%%%%%%%%%%%%%%%%
\begin{figure}[tbp!]
\begin{center}
\includegraphics[width=0.9\columnwidth]{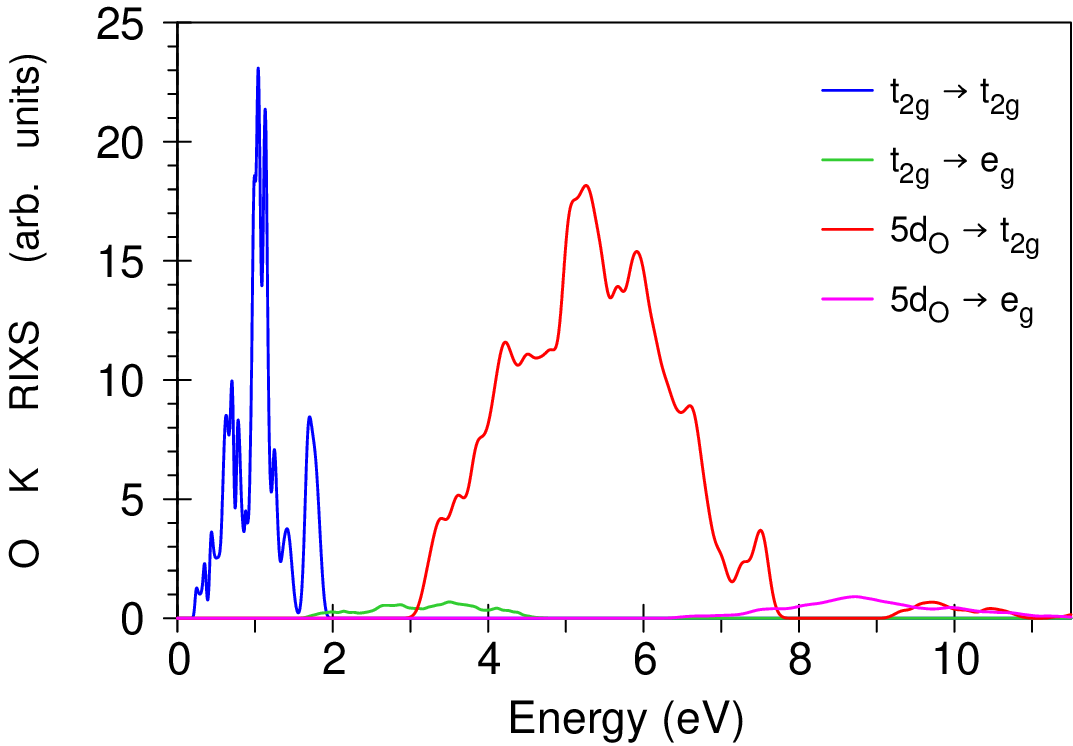}
\end{center}
\caption{\label{rixs_O_K_BIIO}(Color online) The theoretically
  calculated resonance inelastic x-ray scattering (RIXS) spectrum at
  the O $K$ edge in Ba$_3$InIr$_2$O$_9$ in the GGA+SO+$U$ approach
  ($U_{\rm{eff}}$=2.3 eV). }
\end{figure}
%%%%%%%%%%%%%%%%%%%%%%%%

Figure \ref{rixs_O_K_BIIO} shows the theoretically calculated RIXS spectrum at
the O $K$ edge in Ba$_3$InIr$_2$O$_9$ in the GGA+SO+$U$ approach. We found
that the $\tg \rightarrow \tg$ transitions (the blue curve), which are situated 
below 2 eV, are quite intensive and even exceed the 5$d_{\rm{O}}$ $\rightarrow \tg$
transitions (the red curve). This is quite unusual because the latter transitions
in some other transition metal oxides, such as Sr$_2$IrO$_4$ \cite{AKB24} and
Ca$_3$Ru$_2$O$_7$ \cite{AKB24b}, are dominant. The $\tg \rightarrow \eg$
(the green curve) and 5$d_{\rm{O}}$ $\rightarrow \eg$ transitions (the magenta curve)
are rather small. Experimental measurements of the RIXS spectrum at the O
$K$ edge in Ba$_3$InIr$_2$O$_9$ are highly desirable.

\section{Conclusions}

To summarize, we have investigated the electronic and magnetic
structures of Ba$_3$InIr$_2$O$_9$ in the frame of the fully
relativistic spin-polarized Dirac approach. We have also presented
comprehensive theoretical calculations of the RIXS spectra at the Ir
$L_3$ and oxygen $K$ edges.

The delicate interplay between electron correlations, SOC, intersite
hopping, and a crystal field splitting leads to a strongly competing
ground state for Ba$_3$InIr$_2$O$_9$. We found that the ground
magnetic state in Ba$_3$InIr$_2$O$_9$ is AFM$_2$ with the FM
intra-dimer and AFM inter-dimer coupling. The GGA and GGA+SO
approaches give a metallic ground state in Ba$_3$InIr$_2$O$_9$. It is
in contradiction with electric conductivity and energy loss
measurements, which firmly indicate an insulating character of
Ba$_3$InIr$_2$O$_9$. After taking into account the Coulomb
correlations in the frame of the GGA+SO+$U$ method, we obtained the
energy gap for $U_{\rm{eff}}$ = 1.5 eV for the AFM ground state. The
calculations show that the energy gap is already formed in the GGA+$U$
approximation without SOC.  Ba$_3$InIr$_2$O$_9$ can be classified as a
Mott insulator, since it was expected to be metallic from GGA band
structure calculations. We found that the best agreement between the
calculated and experimentally measured RIXS spectra at the Ir $L_3$
edge in Ba$_3$InIr$_2$O$_9$ can be achieved for $U_{\rm{eff}}$ = 2.3
eV.

Ba$_3$InIr$_2$O$_9$ contains two formula units and two structural
Ir$_2$O$_9$ dimers in the unit cell, which are IrO$_6$ face-shared
octahedra connected via O-In-O paths along the $c$ axis. The Ir-Ir
distance within the dimer is quite small and equal to 2.596\AA\, at
low temperature. As a result, there is clear formation of the $a_{1g}$
molecular orbital in Ba$_3$InIr$_2$O$_9$.

The theoretically calculated Ir $L_3$ RIXS spectrum is in good
agreement with the experiment. We found that the low energy part of
the RIXS spectrum $\le$2 eV corresponds to intra-{\tg} excitations. It
has quite a rich fine structure with at least twelve well separated
peaks. Such a shape of the RIXS spectrum can be explained by a
specific energy band structure of Ir {\tg} states with nine DOS peaks
below $E_F$ and three empty narrow DOS peaks separated by energy gaps.
The interband transitions between these nine occupied and three empty
bands produce the rich structure of the Ir $L_3$ RIXS spectrum below 2
eV. The appearance of multiple peaks in the RIXS spectrum is a direct
consequence of strong noncubic crystal field splitting originating
from the distorted octahedral environment of the Ir ions. The RIXS
peak located at $\sim$4 eV was found to be due to $\tg \rightarrow
\eg$ transitions with some additional contributions from 5$d_{\rm{O}}$
$\rightarrow \tg$ transitions. The strong fine structure at $\sim$9 eV
is due to 5$d_{\rm{O}}$ $\rightarrow$ {\eg} transitions.

We found rather small momentum dependence of the excitations in
Ba$_3$InIr$_2$O$_9$, which is typical for 5$d$ transition oxides. Our
investigation of the Ir $L_3$ RIXS spectrum as a function of incident
photon energy $E_i$ shows that the intra-{\tg} excitations steadily
decrease with increasing the incident photon energy, whereas the high
energy peaks corresponding to the $\tg \rightarrow \eg$ transitions
and the charge transfer transitions 5$d_{\rm{O}}$ $\rightarrow \tg$
possess oscillation dependence on the incident photon energy.

\section*{Acknowledgments}

We are thankful to Dr. Alexander Yaresko from the Max Planck Institute
FKF in Stuttgart and Dr. Yuri Kucherenko from the G. V. Kurdyumov
Institute for Metal Physics of the N.A.S. of Ukraine for helpful
discussions. The studies were supported by the National Academy of
Sciences of Ukraine within the budget program KPKBK 6541230 "Support
for the development of priority areas of scientic research".

%\bibliography{./jprb,./book,./DP}

\newcommand{\noopsort}[1]{} \newcommand{\printfirst}[2]{#1}
  \newcommand{\singleletter}[1]{#1} \newcommand{\switchargs}[2]{#2#1}

\end{document}